# FP-tree and COFI Based Approach for Mining of Multiple Level Association Rules in Large Databases

**Virendra Kumar Shrivastava**
Department of Computer Engineering,
Singhania University,
Pacheri Bari, (Rajsthan), India
vk_shrivastava@yahoo.com

**Dr. Parveen Kumar**
Department of Computer Science & Engineering,
Asia Pacific Institute of Information Technology,
Panipat (Haryana), India
drparveen@apiit.edu.in

**Dr. K. R. Pardasani**
Dept. of Maths & MCA,
Maulana Azad National Inst. Of Tech.,
Bhopal, (M. P.) India
kamalrajp@hotmail.com

## Abstract

*In recent years, discovery of association rules among itemsets in a large database has been described as an important database-mining problem. The problem of discovering association rules has received considerable research attention and several algorithms for mining frequent itemsets have been developed. Many algorithms have been proposed to discover rules at single concept level. However, mining association rules at multiple concept levels may lead to the discovery of more specific and concrete knowledge from data. The discovery of multiple level association rules is very much useful in many applications. In most of the studies for multiple level association rule mining, the database is scanned repeatedly which affects the efficiency of mining process. In this research paper, a new method for discovering multilevel association rules is proposed. It is based on FP-tree structure and uses co-occurrence frequent item tree to find frequent items in multilevel concept hierarchy.*

## 1. Introduction

Association Analysis [1, 2, 5, 11] is the discovery of association rules attribute-value conditions that occur frequently together in a given data set. Association analysis is widely used for market basket or transaction data analysis.

Association Rule mining techniques can be used to discover unknown or hidden correlation between items found in the database of transactions. An association rule [1,3,4,7] is a rule, which implies certain association relationships among a set of objects (such as 'occurs together' or 'one implies to other') in a database. Discovery of association rules can help in business decision making, planning marketing strategies etc.

Apriori was proposed by Agrawal and Srikant in 1994. It is also called the level-wise algorithm. It is the most popular and influent algorithm to find all the frequent sets.

The mining of multilevel association is involving items at different level of abstraction. For many applications, it is difficult to find strong association among data items at low or primitive level of abstraction due to the sparsity of data in multilevel dimension. Strong associations discovered at higher levels may represent common sense knowledge. For example, instead of discovering 70% customers of a supermarket that buy milk may also buy bread. It is also interesting to know that 60% customer of a super market buys white bread if they buy skimmed milk. The association relationship in the second statement is expressed at lower level but it conveys more specific and concrete information than that in the first one.

To describe multilevel association rule mining, there is a requirement to find frequent items at multiple level of abstraction and find efficient method for generating association rules. The first requirement can be full filled by providing concept taxonomies from the primitive level concepts to higher level. There are possible to way to explore efficient discovery of multiple level association rules. One way is to apply the existing single level association rule mining method to mine multilevel association rules. If we apply same minimum support and minimum confidence thresholds (as single level) to the multiple levels, it may lead to some undesirable results. For example, if we apply Apriori algorithm [1] to find data items at multiple level of abstraction under the same minimum support and minimum confidence thresholds. It may lead to generation of some uninteresting associations at higher or intermediate levels.

1. Large support is more likely to exist at high concept level such as bread and butter rather than at low concept levels, such as a particular brand of bread and butter. Therefore, if we want to find strong relationship at relatively low level in hierarchy, the minimum support threshold must be reduced substantially. However, it may lead to generation of many uninteresting associations such as butter => toy. On the other hand it will generate many strong association rules at a primitive concept level. In order to remove the uninteresting rules generated in association mining process, one should apply different minimum support to different concept levels. Some algorithms are developed and progressively reducing the minimum support threshold at different level of abstraction is one of the approaches [6,8,9,14].

This paper is organized as follows. The section two describes the basic concepts related to the multiple level association rules. In section three, a new method for mining the frequent pattern at multiple-level is proposed. Section four describes the conclusions of proposed research work.

## 2. Multiple-level Association Rules:

To study the mining of association rules from a large set of transaction data, let assume that the database contains:
  i. an item data set which contains the description of each item in I in the form of <$A_i$, description$_i$>, where $A_i$ Є I, and
  ii. a transaction data set T , which consists of a set of transactions <$T_i$, {$A_p$, . . .,$A_q$}>, where $T_i$ is a transaction identifier and $A_i$ Є I (for i . p, . . . , q).

**2.1 Definition:** A pattern or an itemset A, is one item $A_i$ or a set of conjunctive items $A_i$ ^…^ $A_j$, where $A_i$, . . . , $A_j$ Є I. The support of a pattern A in a set S, s(A/S), is the number of transactions (in S) which contain A versus the total number of transactions in S. The confidence of A => B in S, c(A => B/S), is the ratio of s(A ^ B/S) versus s(A/S), i.e., the probability that pattern B occurs in S when pattern A occurs in S.

To generate relatively frequent occurring patterns and reasonably strong rule implications, one may specify two thresholds: minimum support s´, and minimum confidence c´. Observe that, for finding multiple-level association rules, different minimum support and/or minimum confidence can be specified at different levels.

2.2 Definition: A patterns A is frequent in set S at level l if the support of A is no less than its corresponding minimum support threshold s´. A rule "A => B/S" is strong if, for a set S, each ancestor (i.e., the corresponding high-level item) of every item in A and B, if any, is frequent at its corresponding level, "A ^ B/S" is frequent (at the current level), and the confidence of "A => B/S" is no less than minimum confidence threshold at the current level.

Definition 2.2 implies a filtering process which confines the patterns to be examined at lower levels to be only those with large supports at their corresponding high levels. Therefore, it avoids the generation of many meaningless combinations formed by the descendants of the infrequent patterns. For example, in a sales_transaction data set, if milk is a frequent pattern, its lower level patterns, such as fat free milk, will be examined; whereas if fruit is an infrequent pattern, its descendants, such as orange, will not be examined further.

Example 2.1
To find multiple-level strong associations in the database in Table 1, for the purchase patterns related to category, content and brand of the foods.

Table 1
A sales_transacton database

| TID | Barcode |
|---|---|
| 1001 | 20005, 40001, 50022, 60034, 60045,… |
| 1002 | 20001, 50022, 50023, …. |

Table 2
A sales_item Table

| barcode | Category | brand | content | … | .. | .. |
|---|---|---|---|---|---|---|
| 20005 | Milk | oganic | Fat free | | | |
| ….. | … | ……. | ….. | …. | … | .. |

Table 3
Generalized sales_item description Table

| GID | Barcode | category | brand | content |
|---|---|---|---|---|
| 101 | {20005, …….} | milk | oganic | Fat free |
| .. | ….. | ….. | ….. | ……. |

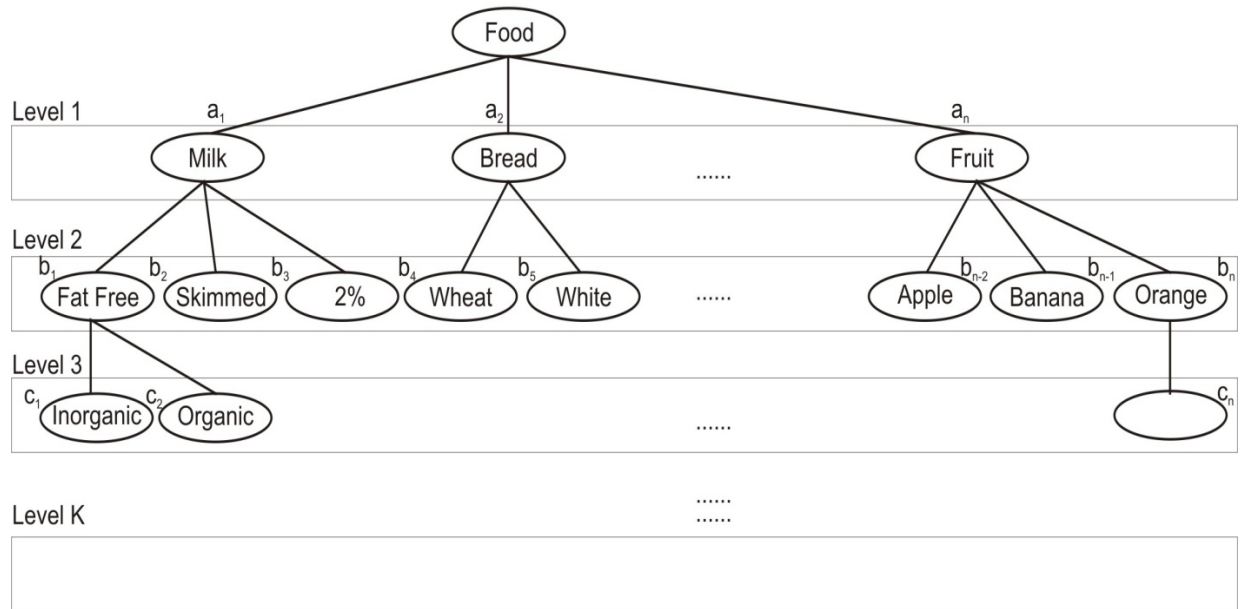

Figure 1: concept hierarchy

We can obtain the relevant part of the sales_item description from table 2 and generalized into the generalized sales_item description table 3. For example, the tuples with the same category, brand and content in table 2 are merged into one, with their barcodes replaced by the barcode set. Each group is treated as an atomic item in the discovery of the lowest level association rules. For example, the association rules discovered related to milk will be only in relevance to (at the low concept levels) brand (such as fat free) and content (such as organic) but not size, etc.

The Table 3 describes a concept tree as given in figure 1. The taxonomy information is given in the Table 3. Let assume category (such as "milk") represent the first level concept, content (such as "fat free") for the second level one, and brand (such as "organic") for the third level one.

In order to discover association rules, first find large patterns and strong association rules at the top most concept level. Let the minimum support at this level be 6% and the minimum confidence be 55% . We can find a large 1-item set with support in parentheses (such as "milk (20%), Bread (30%), fruit (35%)" , a large 2-item set etc. and a set of strong association rules (such as "milk => fat free (60%)")  etc.

At the second level, only the transactions which contain the frequent items at the first level are processed. Let the minimum support at this level be 3% and the minimum confidence is 35%. One may find frequent 1-itemsets: wheat bread (15%),. . . and frequent 2-itemsets: wheat bread (6 percent), 2% milk (7%) . . .  and strong association rules:2% milk => wheat bread (55%),. . .etc. The process repeats at even lower concept levels until there is no frequent patterns can be found.

## 3. Proposed Method for Discovering Multilevel Association Rules

In this section, we propose a method for discovering multilevel association rules. This method uses a hierarchy information encoded transaction table instead of the original transaction table. This is because, first a data mining query is usually in relevance to only a portion of the transaction database, such as food, instead of all the items. Thus, it is useful to first collect the relevant set of data and then work repeatedly on the task related set.  Second, encoding can be performed during the collection of task related data and thus there is no extra encoding pass required. Third, an encoded string, which represents a position in a hierarchy, requires lesser bits than the corresponding bar code. Thus, it is often beneficial to use an encoded table. Although our method does not rely on the derivation of such an encoded table because the encoding can always be performed on the fly.

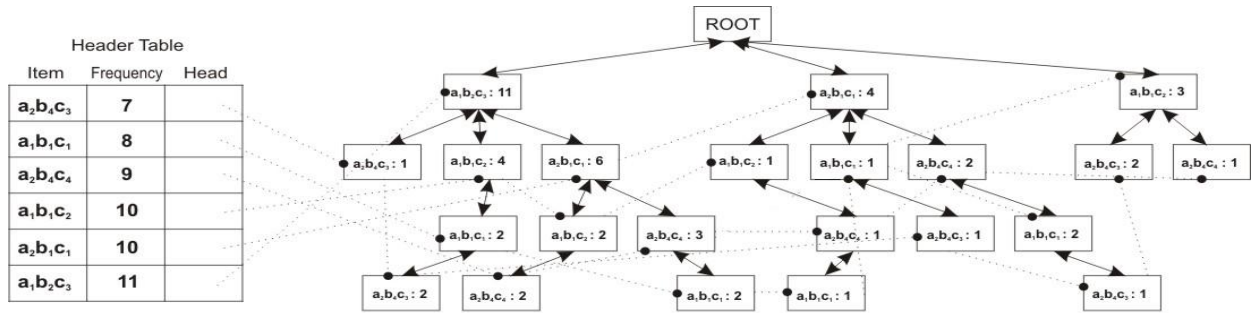

Figure 2. FP(3)-tree (atomic level)

We propose an encoding method which is different from previous one and more general. For example, the item `inorganic fat free milk' is encoded as `$a_1b_1c_1$' in which the first character, `$a_1$', represents `milk' at level-1, the second, `$b_1$', for `Fate free (milk)' at level-2, and the third, `$c_1$', for the brand `inorganic' at level-3. Repeated items (i.e., items with the same encoding) at any level will be treated as one item in one transaction.

The proposed method consists of two main stages. Stage one is the construction of a modified Frequent Pattern tree. Stage two is the repetitive building of small data structures, the actual mining for these data structures, and their release. The association rules are generated at multiple- level using the frequent patters at related concept level.

**Construction of FP-tree**

The FP-tree [10] is created in two phases. In the first phase, we scan the database to generate the ordered list of frequent 1-itemsets. This list is stores in a table called header table, in which the items and their respective support are stored along with pointer to the first occurrence of the item in the FP-tree. The second phase constructs the PF-tree. This phase also requires the full scan of the database. For each transaction read, only the set of frequent items present in the header and sort in descending order according to their support. These sorted items are used to construct the FP-tree. For the first item of the sorted dataset, check if it exists as one of the children of the root then increment the support of this node by 1 otherwise, add a new node for this item with support 1 as a child under the root node. Repeat the same procedure for the next item on the sorted itemset. When we add a new item-node to the FP-tree, a link is maintained between this new item-node and its entry in the header table. The header table maintains one pointer per item that points to the first occurrence of this item in the FP-tree structure. Here links between the item-nodes are bi-directional. The bi-directional pointers enable the mining process by making the traversal of tree easier.

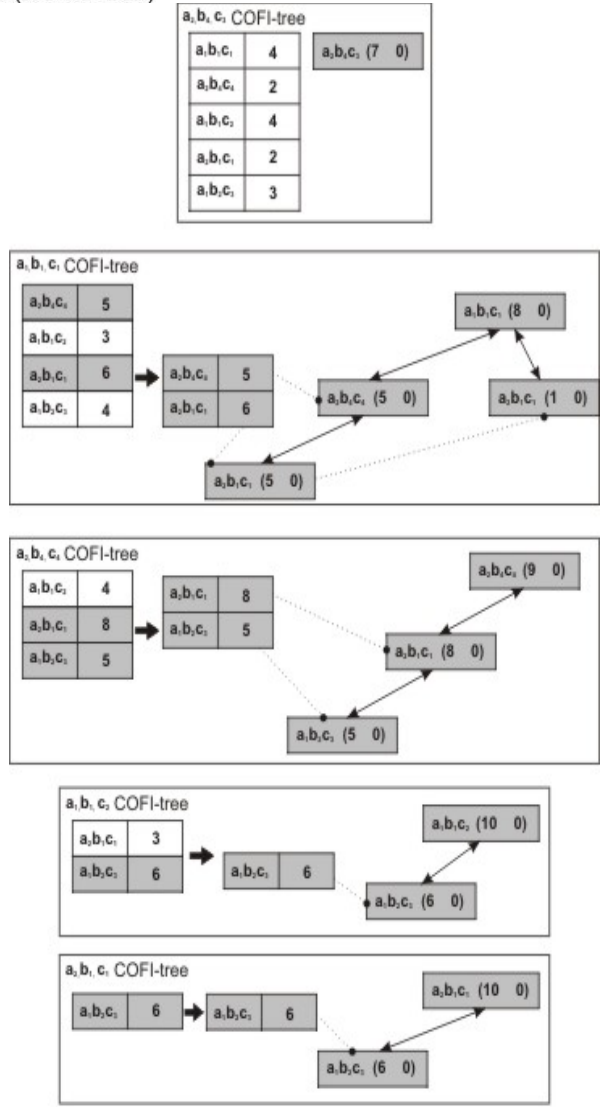

Figure 3. COFI-trees

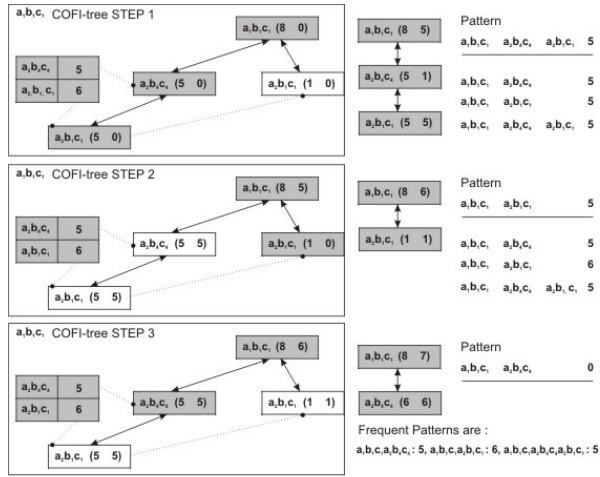

Figure 4, Steps to generate frequent patterns related to item E

**Construction of modified FP(l)-tree**

The FP'-tree is a higher concept FP-tree transformed from its lower level concept tree. We use FP(l)-tree to denotes the FP'-tree at the concept level l. We used algorithm [8,10] for constructing FP(l)-tree given the FP-tree of atomic level and support threshold of level l. In step one, the form of items in header table as well as nodes in FP-tree is changed to that of level l by replacing the encoded character which represent the lower levels with "*". The second step is for each item in the header, if its support does not satisfy given minimum support threshold then remove the item and its relative nodes from the header table and FP(l)-tree respectively. For each item in the new header table, merge the identical ones and relative nodes in the FP(*l*)-tree. (a) Remove the recurrent items and relative nodes, and cumulate the support counts to keep ones respectively.(b) Sort the items in the header table and nodes in the FP(*l*)-tree in ascending order.(c) Adjust the node-links and path-links in the FP (*l*)-tree.

**Co-Occurrence of Frequent-Item-Tree**

Our method for computing frequencies relies on creating independent, relatively small trees for each frequent item in the header table of the FP(l)-Tree called COFI trees [12,13]. Pruning is done by removing all non frequent items with respect to the main frequent item of tested COFI-tree. We are using anti monotone property called global frequent / local non frequent property [12]. It is similar to the Apriory property [1]. It eliminates at $i^{th}$ level all non frequent items that will not participate in (i+1) level of candidate item generation. To eliminate frequent items which are in the i-itemset and it is sure that they will not take part in the (i+1) candidate set. This is used to find all frequent patterns with respect to one frequent item, which is the base item of the tested COFI-tree. As we know that all items that participate in the construction of the COFI-tree are frequent with respect to the global database, but it does not imply that they are also locally frequent with respect to the based item in the COFI-tree.

The small COFI-trees are similar to the FP-tree [10]. However, the COFI-trees have bi-directional links in the tree allowing bottom up scanning as well. The nodes in the COFI-tree contain item label, frequency counter and a contribution counter. The contribution counter cumulates the participation of the item in all patterns already discovered in the current COFI-tree. The difference between the contribution in the node and the contribution in the header is that the counter in the node counts the participation of the node item in all paths where the node appears, while the new counter in the COFI-tree header counts the participation of the item globally in the tree. The COFI-tree for a given frequent item x contains only nodes labeled with items that are more frequent of as frequent as x.

**Algorithm 3.1 COFI:** Creating with pruning and Mining
COFI-trees for FP(l)-tree
**Input:** modified FP(l)-Tree, the support threshold *s* of level *l*
**Output:** frequent item sets
**Method:**
1. A= the least frequent item on the header table of FP(l)-Tree
2. While (There are still frequent items) do
2.1 add up the frequency of all items that share item (A) a path. Frequencies of all items that share the same path are the same as of the frequency of the (A) items
2.2 Eliminate all non-locally frequent items for the frequent list of item (A)
2.3 Create a root node for the (A)-COFI-tree with both *frequency-count* and *contribution-count* = 0
2.3.1 C is the path of locally frequent items in the path of item A to the root
2.3.2 Items on C form a prefix of the (A)-COFI-tree.
2.3.3 If the prefix is new then Set *frequency-count*= frequency of (A) node and *contributioncount*= 0 for all nodes in the path
Else
2.3.4 Update the *frequency-count* of the already exist part of the path.
2.3.5 Update the pointers of the *Header list* if needed
2.3.6 find the next node for item A in the FP(l)-tree and go to 2.3.1
2.4 MineCOFI-tree (A)
2.5 Release (A) COFI-tree

2.6 A = next frequent item from the header table of FP(l)-Tree
3. Goto 2

**Function:** MineCOFI-tree (A)
1. nodeA = select next node //Selection of nodes starts with the node of most locally frequent item and following its chain, then the next less frequent item with its chain, until we reach the least frequent item in the *Header list* of the (A)-COFI-tree
2. while (there are still nodes) do
2.1 D = set of nodes from nodeA to the root
2.2 F = nodeA.*frequency-count*-nodeA. *contribution-count*
2.3 Generate all Candidate patterns X from items in D. Patterns that do not have A will be discarded.
2.4 Patterns in X that do not exist in the A-Candidate List will be added to it with frequency = F otherwise just increment their frequency with F
2.5 Increment the value of *contribution-count* by F for all items in D
2.6 nodeA = select next node
3. Goto 2
4. Based on support threshold s remove non-frequent patterns from A Candidate List.

**Algorithm 3.2**
**Input:** Candidate rule-set $R_1$, FP(L)-tree for each concept level, support threshold *s* and confidence *c* at each concept level.
$A_i$ - the antecedent of rule $r_i \in R_1$ represents;
*n* - the total transactions;
$i_m$ - the item which has the lowest concept level among the items in $r_i$;
*FP(l)-tree* - the corresponding FP-tree of the concept level of $i_m$.
**Output:** The confirmed rule-set $R_1$.

**Method:**
For each rule $r_i$, if its support and confidence are NULL, calculate its support and confidence by following steps:
(a) Start from the head (in header table) of $i_m$, and follow its node-links and located paths in FP(*l*)-tree to find all other items, which belong to the lower concept levels of the items in $r_i$
(b) Calculate the support counts of $r_i$ and $A_i$ with the COFI-tree and item $i_m$ derives, and sum them respectively to get the support count $s_i$ of $r_i$ and the support count $s_i'$ of $A_i$.

(c) If $s_i/n s$ and $s_i/s_i' \geq c$, then keep the rule $r_i$ in the $R_1$ and delete the corresponding rules which have the same group ID if they are atomic rules; else delete $r_i$ from the $R_1$.

### 4. Conclusion

In this research works, we have proposed a generalized encoding method and combining FP growth tree with COFI for mining of multilevel association rules from large database. This proposed approach uses FP(l)-tree to construct FP-tree for the level l. To find frequent patters this new method creates COFI-tree which reduces the memory usage in comparison to FP-growth effectively in efficient way. Therefore, it can mine larger database with smaller main memory available. This method uses the non recursive mining process and a simple traversal of the COFI-tree, a full set of frequent items can be generated. It also uses an efficient pruning method that is to remove all locally non frequent patters, leaving the COFI-tree with only locally frequent items. It reaps the advantages of both the FP growth and COFI.

## AUTHORS PROFILE

Virendra Kumar Shrivastava has completed his M. Tech. (Computer Technology) from School of IT, RGPV (Sate Technological University of M. P.) Bhopal, India. He is Associate Professor at Asia Pacific Institute of Information Technology SD India Panipat (Haryana) India. His research area is Data Mining. Presently he is perusing Ph. D. in Department of Computer Engineering, Singhania University, Pacheri Bari (Raj.) India.

Dr. Parveen Kumar has obtained Ph. D. in Computer Science from Kurukshetra University, Kurukshetra (Haryana), India. Presently he is working as Professor cum Director Research at Asia Pacific Institute of Information Technology SD India, Panipat (Haryana) india. His research interest includes Check Point and Data Mining.

Dr. Kamal Raj Pardasani is working as Professor and Head in Department of Mathematics and Dean Research and Development Maulana Azad National Institute of Technology, Bhopal. He did his Ph. D. in applied Mathematics in 1988. His current research interests are Computational Biology, Data Mining, Bio-computing and Finite Element Modeling.